%% file: main.tex
\documentclass[letterpaper,twocolumn,10pt]{article}

\usepackage{times}
\usepackage{fullpage}
\usepackage[compact]{titlesec}

\usepackage{amsmath}
\usepackage{graphicx}
\usepackage{setspace}
\usepackage{subfig}
\usepackage[hyphens]{url}
\usepackage{hyperref}
\usepackage{xspace}
\usepackage{flushend}

\long\def\com#1{}

\begin{document}

\setlength{\pdfpageheight}{\paperheight}
\setlength{\pdfpagewidth}{\paperwidth}

\hyphenation{Nym-OS}
\hyphenation{Anon-VM}
\hyphenation{Comm-VM}

\newcommand{\winon}{Nymix\xspace}
\newcommand{\Nymvm}{NymBox\xspace}
\newcommand{\Nymvms}{NymBoxes\xspace}
\newcommand{\nymvm}{nymbox\xspace}
\newcommand{\nymvms}{nymboxes\xspace}

\title{\vspace{-1em}%
	Managing \Nymvms for Identity and Tracking Protection%
	\vspace{-2em}}

\date{}
\maketitle

\thispagestyle{empty}

\begin{abstract}
\input{abs}
\end{abstract}
\input{intro}
\input{motiv}
\input{arch}
\input{impl}
\input{eval}
\input{rel}
\input{disc}
\input{conc}



\bibliographystyle{abbrv}
\bibliography{theory,net,os,soc,sec}

\end{document}

%% file: abs.tex
Despite the attempts of well-designed anonymous communication tools
to protect users from tracking or identification,
flaws in surrounding software (such as web browsers)
and mistakes in configuration may leak the user's identity.
We introduce \winon,
an anonymity-centric operating system architecture
designed ``top-to-bottom'' to strengthen identity- and tracking-protection.
\winon's core contribution is OS support for {\em nym-browsing}:
independent, parallel, and ephemeral web sessions.
Each web session, or pseudonym, runs
in a unique virtual machine (VM) instance
evolving from a common base state
with support for long-lived sessions
which can be anonymously stored to the cloud,
avoiding de-anonymization despite potential confiscation or theft.
\winon allows a user to safely browse the Web
using various different transports simultaneously
through a pluggable communication model
that supports Tor, Dissent, and a private browsing mode.
In evaluations, \winon consumes 600 MB per \nymvm
and loads within 15 to 25 seconds.

%% file: intro.tex
\section{Introduction}

Today's Internet users must increasingly assume
that by default all of their online activities are tracked
and that detailed profiles of their identities and behaviors
are being collected by every Web site they visit~\cite{soltani09flash},
sold for marketing purposes~\cite{duhigg12how,oecd13exploring},
and ingested into mass surveillance systems~\cite{risen13nsa}.
Users may wish to protect their online activities
from being tracked or linked with their real identities, however,
or to access the Internet under several distinct
{\em roles}, {\em personas}, or {\em pseudonyms}.
Anonymity and pseudonymity are desirable to many types of users,
from repressed minorities~\cite{stein03queers} and
dissidents in authoritarian countries~\cite{lim12clicks,howard13democracy},
to women wishing to hide their pregnancies
from advertisers~\cite{hill12target,hill14pregnancy},
to celebrity authors desiring
``feedback under a different name''~\cite{watts13rowling}.

Anonymity protocols such as Tor~\cite{dingledine04tor},
Dissent~\cite{wolinsky12scalable},
and Aqua~\cite{leblond13anon}
obscure a user's network location,
but {\em client-side} weaknesses can break this anonymity.
Web sites may still be able to track the user
via third-party plug-ins that circumvent the anonymous
channel~\cite{perry11firefox,fleischer12tor},
via browser fingerprints~\cite{eckersley10browser},
employ software exploits
to ``stain'' the client for long-term tracking~\cite{mullenize},
or de-anonymize users
directly~\cite{goodin13attackers,schneier13attacking}.
Anonymity-oriented Linux distributions such as
Tails~\cite{tails} and Whonix~\cite{whonix} mitigate some risks,
but leave to the user the error-prone task of
managing different online roles or pseudonyms.
Users can accidentally de-anonymize themselves
by logging in to a sensitive account from the wrong browser window
or otherwise neglecting to protect anonymity
even once~\cite{erratasec12lulzsec},
or by posting a photo without realizing that the JPEG may contain
GPS coordinates~\cite{oakes12hacking}.

\begin{figure}[t]
\includegraphics[width=0.49\textwidth,trim=8 0 0 0,clip]{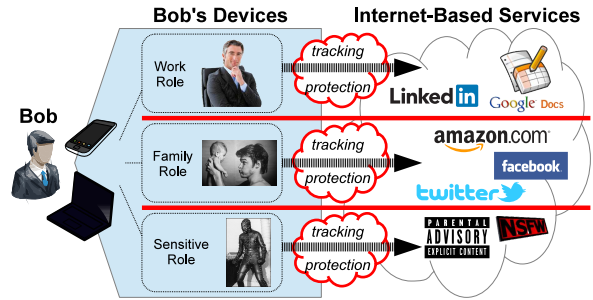}
\caption{\winon is a client OS architecture designed to enable users
	to manage multiple {\em roles} in their online life,
	and offer strong protections against their roles being
	tracked or linked.}
\label{fig:roles}
\end{figure}

To address this need we present \winon,
the first OS architecture designed
to help users manage online roles, or {\em nyms},
and to protect these nyms systematically
from accidental or malicious linking.
As illustrated in Figure~\ref{fig:roles},
\winon aims to offer {\em end-to-end} isolation between nyms,
separating all client-side state and browser activity related to each nym
into protected virtual machines, or {\em \nymvms}.
\winon connects these \nymvms to the Internet
{\em only} via separate instances of
network tracking protection systems, such as Tor,
protecting nyms from being linked by the online services they are used to access.

\winon enables and encourages users
to create ephemeral, ``throwaway'' \nymvms on demand
for activities requiring no long-term state,
such as reading news,
reducing the user's vulnerability
to long-term tracking or intersection attacks~\cite{kedogan02limits}.
Users can also create {\em persistent \nymvms} when needed,
which can remember long-lived state such as login credentials
for pseudonymous Internet accounts.
Unlike common password managers~\cite{keepass},
\winon maintains and structurally enforces
an explicit binding between each role a user plays online,
the network login credentials related to each role,
and all client-side state such as browser history related to each role.
By binding client state and credentials to \nymvms,
\winon reduces the user's risk of accidentally entering credentials
in the wrong context or browser window --
when using the {\em correct} \nymvm
the user need not enter those credentials at all.

Like Tails~\cite{tails},
\winon can boot from a USB drive for easy deployability
and avoids leaving any history trail on the host machine,
offering deniability in situations where
installing anonymity tools may be dangerous or impossible.
\winon encrypts and saves persistent \nymvm state to either local
media or anonymous cloud storage.
By default, \winon updates nym state {\em only}
at explicit user request (e.g., after the first login),
and not after every browsing session,
to protect the user from staining attacks on the \nymvm's state,
adding further deniability and history protection
if the nym is ever compromised.

\com{

\begin{figure}[t]
\includegraphics[width=0.45\textwidth,trim=8 0 0 0,clip]{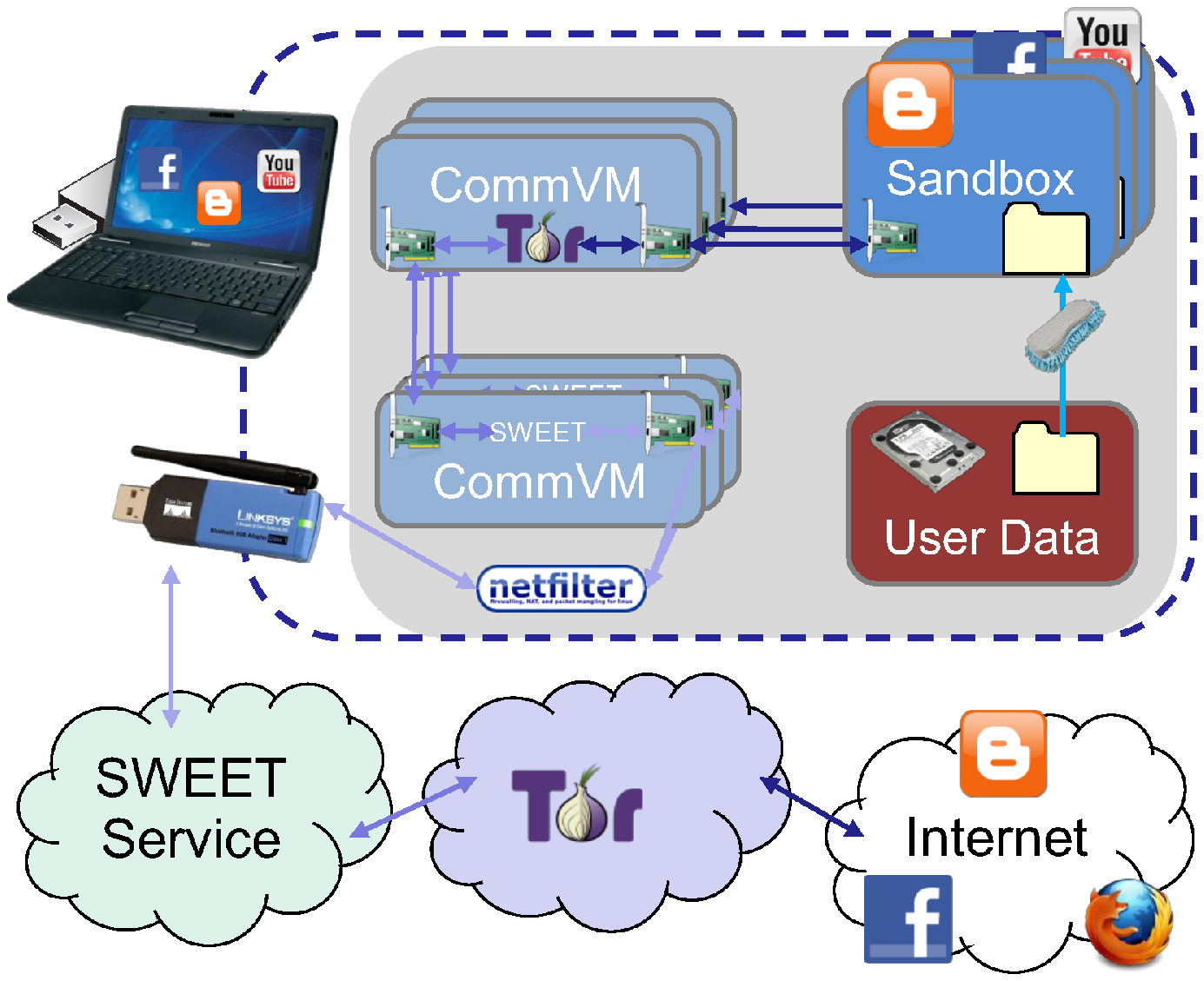}
\caption{\winon organization}
\label{fig:organization}
\end{figure}

Previous work on building anonymous environments
support amnesiac behavior~\cite{tails},
homogeneous environments~\cite{tails,whonix},
enforced anonymous communication~\cite{whonix},
and scrubbing of file types~\cite{aura06scanning,
bier09rules,sweeney96replacing,voisin13mat}.
Amnesiac behavior provides plausible deniability
but requires the user to remember and reproduce state.
Existing environments support a single communication path,
usually through Tor~\cite{dingledine04tor},
limiting their applicablity for complex use,
such as concurrent anonymous and non-anonymous Web browsing.
Current scrubbing efforts require knowledge of the tool,
which unaware users may bypass without a second thought.
Related work on securing web browsing~\cite{tang10trust, mickens11atlantis}
have eliminated many browser-based attacks,
however, such efforts offer no support for anonymous communication
and cannot easily be applied to commodity software or systems.

}

\winon has been under development for the past two years,
predating public knowledge that governments use malware
to track~\cite{mullenize}
and de-anonymize users~\cite{goodin13attackers,schneier13attacking},
two key attack vectors that \winon's design addresses.
Throughout its development, the prototype has undergone
regular design review and adversarial testing by an independent red-team.
The prototype is based on Ubuntu 14.04,
and uses two QEMU/KVM virtual machines for each \nymvm:
one to run communication tools such as Tor,
and the other for the Web browser and associated plugins.
\winon supports multiple pluggable anonymous communication systems
including Tor, Dissent,
and a lightweight {\em incognito mode}
that imposes minimal overhead but
does not protect against network-level tracking.

Our experimental results show that \winon offers similar performance
to running the
software natively or in similar software distributions, such as Tails.
\winon unsurprisingly requires significantly more memory, however, as each
\nymvm runs two virtual machines with all file system writes stored in RAM.

This paper's key contributions are:
(1) an operating system architecture designed to help users keep
local and online state related to different roles isolated
and protected from tracking or linking;
(2) a composable framework supporting pluggable anonymity tools,
(4) anonymous quasi-persistent nym storage either locally or in the cloud,
(3) user-directed sanitization to control information leaks across nyms,
and
(5) the ability to launch the user's installed OS as a nym
for deniability and history protection.

Section~\ref{sec:challenges} identifies
key challenges for anonymity and pseudonymity.
Section~\ref{sec:arch} presents \winon's trust model and architecture.
Section~\ref{sec:impl} describes our prototype implementation
and our experiences with it.
Section~\ref{sec:eval} evaluates how well our architecture and prototype
handle challenges mentioned earlier.
Section~\ref{sec:rel} discusses related work.
Section~\ref{sec:discussion} reflects on other challenges and future work,
and Section~\ref{sec:conc} concludes.

%% file: motiv.tex
\section{Background and Motivation}
\label{sec:challenges}

This section motivates \winon,
and outlines the key challenges it attempts to address,
via two fictional scenarios.

In the People's Republic of Tyrannistan,
the state-controlled ISP monitors all users' traffic
to censor and suppress dissent.
Bob, a Tyrannistani dissident well aware of these dangers,
uses Tor~\cite{dingledine04tor} at night
to organize protests via his
pseudonymous Twitter account~\cite{lim12clicks,howard13democracy}.
Bob spends his days at the state-run newspaper,
using his laptop to grind out grandiloquent paeans to
Glorious Leader Tyrannistanus Rex IV.
Bob may face imprisonment if the censorware
his job requires detects evidence
of unapproved activities on his laptop's hard disk.
Bob therefore runs Tor only from a Tails USB drive~\cite{tails}.

Bob could still be de-anonymized in many ways
unless he is unerringly cautious.
Since Tails (deliberately) forgets all state after each session,
Bob habitually logs into his Twitter account anew each night --
but if by force of habit
he even once accidentally enters his Twitter credentials
while {\em not} running Tails,
he may be caught~\cite{erratasec12lulzsec}.
Bob's computer and his Web activity~\cite{eckersley10browser,panopticlick}
produce unique fingerprints.
Using an intersection attack~\cite{kedogan02limits},
Tyrannistani police can link Bob
to his pseudonymous Twitter account.
Bob might take a photo with his smartphone at a protest
and post it to his Twitter feed via Tor,
not realizing that the EXIF metadata in the photo contains
GPS coordinates and his smartphone's serial number~\cite{oakes12hacking}.
Even if Bob makes {\em no} such mistakes,
the Tyrannistani police might obtain a zero-day exploit,
and use it against Bob's browser
to inject malware onto his running system,
which reports his true IP and MAC address
to the authorities~\cite{goodin13attackers,schneier13attacking}.

To improve his browsing experience,
Bob begins experimenting with Tails' persistent state
that stores passwords, account settings,
and applications on the same USB device as Tails.
Unfortunately for Bob,
this opens him up to new types of staining~\cite{mullenize}
or fingerprinting,
such as the evercookie~\cite{kamkar10evercookie}
that sticks around even if you disable cookies.
In addition, the USB device now becomes
evidence of Bob's {\em misbehavior}.
Tyrannistan police can confiscate the device,
coerce Bob to decrypt its contents,
and then de-anonymize him.

Life is easier for Alice in Freetopia:
she does not feel in any imminent danger,
and is doing nothing she thinks
the Freetopian Fuzz care about.
She has made some personal choices
that she is not ashamed of,
and likes to discuss online in appropriate forums,
but which she imagines her boss and work colleagues
might not understand~\cite{stein03queers}.
She is also concerned that the web sites she visits,
and the ads they present her,
seem to know more about her than her own family does.
She worries that these web sites might
``out'' her unannounced pregnancy by sending a stream of diaper ads
while her family is around~\cite{hill12target}.
She finds that the only way to keep such personal information secret
is using cloak-and-dagger methods that might themselves
raise suspicions of criminal activity~\cite{hill14pregnancy}.
Thus, although Alice does not think anyone is ``after her,''
she would prefer to enforce a strong and inviolate barrier
between her online activities
related to her work, her family and social life,
and her unannounced preparations for motherhood.

%% file: arch.tex
\section{\winon Architecture}
\label{sec:arch}

This section outlines the \winon architecture,
how it binds pseudonyms to client-side and network state
and protects nyms from being linked,
and how \winon addresses
the challenges discussed above in Section~\ref{sec:challenges}.

\subsection{Architecture Overview}

\begin{figure}[t]
\includegraphics[width=0.45\textwidth,trim=0 0 0 0,clip]{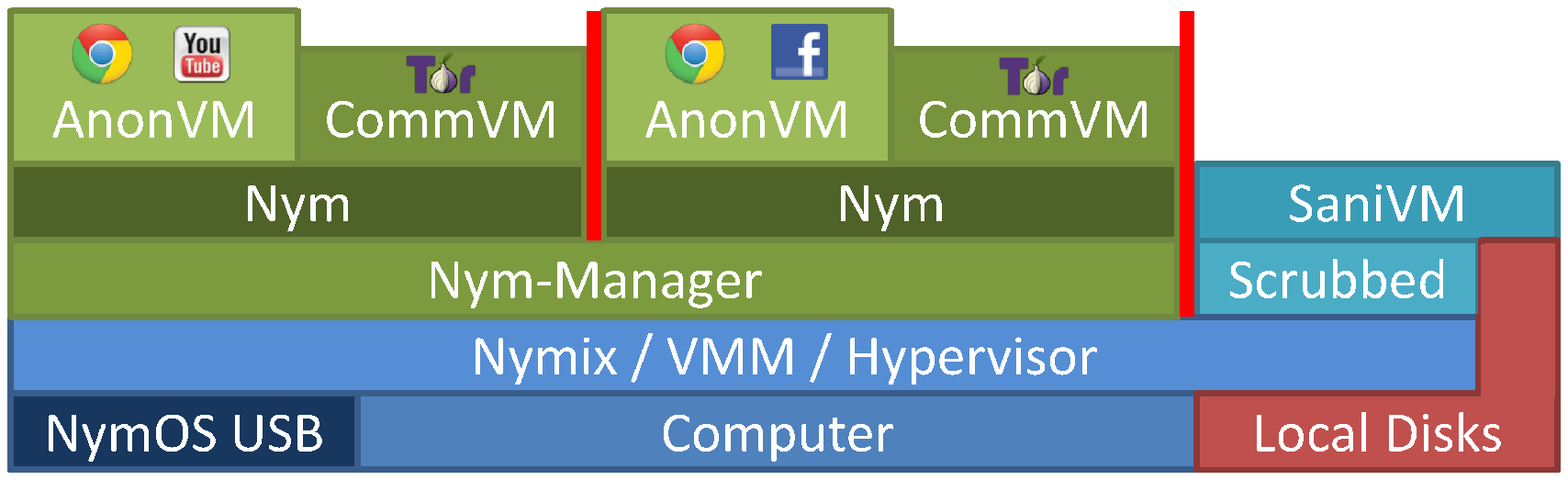}
\caption{Block diagram of \winon architecture.
The hypervisor hosts one or more nyms.
Each nym has an AnonVM for browsing Web content
by communicating through the CommVM, which hosts the anonymizer.
The SaniVM offers sanitized access to local disks
by sharing content scrubbed of personal information to AnonVMs.}
\label{fig:block}
\end{figure}

\winon is designed from the ground up
to offer users strong identity and tracking protection
by giving them explicit, first-class control over {\em pseudonyms}
representing the multiple roles or personas they may use online.
In contrast with the large body of existing work
attempting to improve security or isolation between
distinct users, or between applications run by a single user,
\winon is the first client OS we are aware of
to establish strong {\em separation of roles} through pseudonyms
as a primary OS design objective.
\winon places supervisory control over VM creation, longevity,
and destruction under the control of the user,
binding all client-side application state to a particular nym
via a \nymvm.
To protect the ownership and relationships between different \nymvms,
\winon employs both {\em anonymizers} such as Tor and Dissent
to protect against network-level linkage,
and giving the user a full-featured network client
with the context of each nym,
but deliberately making it difficult for the user to link nyms accidentally
by posting the wrong file or cut-and-paste between the wrong windows.

With \winon, for example,
Alice may instantiate a nym to browse and check her e-mail,
optionally loading the encrypted nym's state anonymously
from a cloud storage system
to avoid leaving a ``footprint'' on her local machine.
During this process,
she wants to check the latest news on Twitter
and instantiates another nym,
which \winon works to keep unlinkable to the first.
Finally, she wishes to post some content
to her pseudonymous blog,
doing that via yet another nym.
She might discard the state of some nyms after each session,
to protect her more sensitive activities from long-term tracking
and intersection attacks~\cite{kedogan02limits},
while preserving the state of other nyms locally or in cloud storage.

As shown in Figure~\ref{fig:block},
\winon's most crucial component is its {\em Nym Manager},
which manages nyms and separates
all client-side browsing and other activities
into separate virtual machines or \nymvms for each nym.
Each \nymvm in fact represents two virtual machines.
All normal client-side activity --
such as running web browsers, their plug-ins,
and other network-connected applications such as mail clients --
is confined to the appropriate nym's {\em AnonVM}.
\winon treats the guest OS and processes in each AnonVM
as untrusted and potentially
compromised~\cite{mullenize,goodin13attackers,schneier13attacking},
and for this reason permits {\em no} interaction between
the AnonVM and the ``outside world'' except
via the nym's corresponding {\em CommVM}.
In this CommVM reside
anonymity and circumvention tools, or {\em anonymizers} such as Tor,
which ensure that interaction between the AnonVM and the Internet
is further protected from network-based tracking.

By isolating each nym's AnonVM from its CommVM,
\winon ensures that software exploits against the AnonVM
cannot compromise anonymity or link nyms without also compromising the VMM.
Launching a separate CommVM for each nym
with an independent instance of the anonymizer, in turn,
ensures that even an anonymizer compromise in one nym
will (hopefully) not compromise other nyms.
Separating CommVMs also ensure that
anonymizer state that is commonly shared or reused for efficiency,
such as Tor circuits,
cannot accidentally reveal the links between different nyms.

\Nymvms have no access to local storage,
i.e., the local hard disk and USB devices.
To safely access personal data,
\winon employs a SaniVM to isolate the user's data
to a single non-networked environment.
The SaniVM sanitizes user data
by either automatic or manual scrubbing
of personally identifiable material from this data
before making it available to an AnonVM.

\subsection{Threat Model}
\label{subsec:threats}

\winon's threat model assumes that
an adversary may be able to compromise software within
a particular AnonVM or CommVM,
install software and inject software exploits,
and even gain root access within the VMs.
However, \winon assumes the adversary cannot access the hypervisor of a
compromised VM nor the host's file systems.
While state-level attackers most likely {\em can} compromise
common VMMs including those \winon uses~\cite{schneier13nsa},
we do not set the unrealistic expectation
of making compromise {\em impossible},
but rather attempt to raise the barrier and cost to attackers
significantly.
We hope eventually \winon could be implemented
in a certified kernel/VMM framework~\cite{klein09sel4,gu11certikos},
further increasing this barrier.

As the CommVM hosts only anonymizing software such as Tor,
an adversary can compromise a CommVM only by attacking an anonymizer directly.
Through a compromised CommVM,
the adversary may learn \winon's public IP address.
While the AnonVM may be compromised through remote exploits,
the CommVM prevents anything in the AnonVM from learning
the user's IP address or other network or physical location information,
so long as the anonymizer in the CommVM is not also compromised.
A compromised AnonVM or CommVM cannot trivially be linked to other
AnonVMs or CommVMs on the same host; however,
attacks may be performed using timing attacks
and side channels~\cite{zhang11cloud,zhang12cloud}.

\winon cannot protect users who obtain
compromised copies of \winon itself,
leaving the important problem of secure software distribution out of scope.
We assume users obtain copies of \winon through trustworthy sources
or verify their authenticity prior to use.
\winon also does not improve the anonymity provided by the anonymizer,
nor can
\winon prevent a user from explicitly de-anonymizing themselves
by typing their real name into an AnonVM for example.

\subsection{Anonymizers and CommVMs}

Anonymizers such as Tor
typically act as client-side Web proxies
that redirect TCP connections through relays to hide their source.
If the Web browser connected to this client-side proxy
is misconfigured or vulnerable, however,
an adversary can exploit that vulnerability to bypass the anonymizer:
for example by invoking a plug-in
that fails to respect
the browser's proxy settings~\cite{perry11firefox,fleischer12tor},
or by using malware to directly read and report
the user's IP address~\cite{goodin13attackers,schneier13attacking}.
Like Whonix~\cite{whonix},
\winon separates anonymizers
from the user's potentially vulnerable Web browsing environment
via two separate virtual machines --
an AnonVM and a CommVM, respectively.
Unlike Whonix, which provides only a static user-managed pair of VM images,
\winon's Nym Manager dynamically launches and manages AnonVMs and CommVMs
and manages their state as part of a user-controlled \nymvm.

The user operates a nym primarily via the AnonVM,
in which the web browser and other applications such as E-mail clients run.
Each AnonVM has a single virtual network link
that connects directly, and {\em only} to,
a CommVM,
which runs an instance of the anonymizer for this nym.
The CommVM redirects all AnonVM traffic to the anonymizer,
which in turns transmits traffic through the anonymity
network via the CommVM's NAT-based Internet connection.
No software in the AnonVM ever gets access to the physical host machine's
IP address, MAC address, or other physical devices
or their trackable device identifiers.

\paragraph{Alternative Anonymizers:}
\winon treats the anonymizer as a pluggable module,
and offers the user a choice of several alternative anonymizers
pre-configured to address different security/performance tradeoffs.
A lightweight {\em incognito mode} uses
simple VPN relaying to provide low-cost anonymization with weak security.
For more sensitive activities the user can employ Tor,
which offers excellent scalability
and good security against moderate adversaries.
Finally, \winon experimentally supports anonymous browsing
via Dissent~\cite{wolinsky12scalable},
an anonymizer based on DC-nets~\cite{chaum88dining}
that in principle offers formally provable traffic analysis resistance
and systematic protection against
intersection attacks~\cite{wolinsky13buddies},
but is less mature and currently less scalable than Tor.
In principle, anonymizers can be combined
by connecting CommVMs in serial, or within the same CommVM:
we have built experimental \winon configurations combining Tor and Dissent
to achieve ``best of both worlds'' anonymity, for example.

While many modern browsers offer incognito or private browsing modes
that promise to erase cookies, history, and other state after a session,
a single state management bug or security vulnerability in the browser
can nevertheless render the user trackable~\cite{aggrawal10private}.
Even in Whonix, such a state management bug --
or malware-based stain attack~\cite{mullenize} --
renders the statically administered browser VM permanently trackable,
and hence vulnerable to long-term intersection attacks,
unless the user manually reinstalls Whonix or resets it
to pristine VM images.
By isolating both the browser and any such stains
in a dynamically managed AnonVM as part of an ephemeral-by-default nym,
\winon ensures that trackable stains disappear immediately
when the nym does.

\subsection{Creating and Configuring \Nymvms}

One of \winon's goals is to be small enough for users
to download conveniently and run from a typical USB drive,
like Tails, 
to support users who wish to leave no trace of their
sensitive Internet access activities on their computers.
A key practical challenge \winon's VM-centric design presents, however,
is that we effectively need to fit the equivalent
of at least three different VM images on the same USB drive:
one containing the host OS atop which \winon is built
(currently Ubuntu Linux),
the second containing an initial disk image for AnonVMs to use
(containing the web browser and other applications),
and the third containing an initial disk image for CommVMs to use
(containing Tor or other anonymizers).
Supporting multiple alternative anonymizers as discussed above
might further increase the number of VM images
that \winon would need to ``ship with.''

To address this challenge,
\winon uses the OS image installed on the \winon USB
as the host OS from which the hypervisor/VMM boots,
as well as the basic VM image for all AnonVMs and CommVMs.
To differentiate these OS images to serve their distinct roles at runtime,
\winon employs union file systems,
which logically stack multiple file systems together
while merging their contents.
The union file system responds to file read accesses
with the contents of that file as it exists in the top most stack.
The file system stores writes into the top most
read-write layer,
shielding lower layers from write access
using copy-on-write.

Live-bootable operating systems such as Tails often
use union file systems
with the top layer using a temporary file system
that resides in RAM.
\winon inserts between the base image and the temporary file system
an additional, intermediary {\em configuration file system}
containing the configuration necessary to start the particular VM --
e.g., one configuration file system for its standard AnonVM configuration,
and a separate configuration file system for the CommVM
representing each alternative anonymizer \winon supports.
The changes this configuration file system makes
include the network configuration files,
the local startup script (/etc/rc.local),
and the window manager startup script.

A \nymvm's temporary file systems store all writes to the file system in RAM.
As a result, turning off a pseudonym results in {\em amnesia} --
\winon wipes any traces that the pseudonym ever existed and
securely erases the AnonVM's and CommVM's memory
immediately on shutting down a pseudonym.
The USB device used during a \winon session remains unchanged,
ensuring that even if confiscated and thoroughly analyzed neither the
computer nor the USB device harbors evidence of \winon use.

After terminating a nym,
\winon removes all state of that nym from memory.
As designed, \winon
and all other existing production solutions retain
traces of that state until reboot;
however, because the hypervisor cannot be accessed without
live confiscation,
such state is likely to be inaccessible by an adversary.
Recent work by Dunn~\cite{dunn12eternal}
explores how much information remains on a host
after a virtual machine has shut down,
yet the hypervisor remains active,
as well as various methods for eliminating it.
\winon could employ these methodologies
to address adversaries with physical access;
however, many of these features require specialized hardware
and additional computational overhead,
so for now we assume that erasing AnonVM and CommVM memory
after shutdown are sufficient.

One security concern, created by
\winon's reuse of the host OS partition as AnonVM and CommVM images,
is that \winon must ensure that the host OS partition
is {\em always} mounted read-only and never modified for any reason.
This implies that any state the user wishes to persist across boots --
such as persistent nyms, as described below --
must be stored elsewhere,
either on different local disks or USB drives
or in cloud storage.
If \winon ever permitted the host OS partition to be modified
from its standard ``distribution'' state,
those modifications, however minute
(even mount-time or access-time updates)
would manifest in the initial states of all AnonVMs subsequently created,
potentially offering adversaries a way to track the user.
While \winon by construction ensures that its host partition
is only ever mounted read-only,
it cannot prevent {\em other} operating systems
from mounting the partition read/write and potentially modifying it
while the USB drive is plugged in.
Although not yet implemented,
we intend to address this risk
by adding a mechanism to check all disk blocks loaded
from the host OS partition into an AnonVM or CommVM
against a well-known Merkle tree~\cite{gassend03merkle} as they are accessed,
and safely shut down rather than risk vulnerability
if a modified block is detected.


\subsection{Quasi-Persistent Nyms}
\label{ss:persistent}

Although ideal from a tracking resistance perspective,
a {\em pure} amnesiac system
that never maintains persistent state across reboots
would inhibit usability,
effectively requiring users to re-initialize
all browser configuration preferences during each session,
and to re-enter login credentials for any pseudonymous Internet accounts
the user might wish to access during the session.
Worse, client OS amnesia can {\em reduce} the security
of users who regularly connect to pseudonymous accounts
such as Alice's dissident Twitter feed in Tyrannistan,
in at least two ways.
First, because Alice must enter
her pseudonymous Twitter username and password during each session,
this procedure is likely to become habit --
but if she ever {\em even once} accidentally performs this procedure
outside of an anonymity-protected context
(e.g., on some other Ubuntu distribution she may sometimes run
with a GUI look-and-feel similar to Tails),
she may be compromised~\cite{erratasec12lulzsec, leger13silkroad}.
Second, state-of-the-art anonymizers like Tor
are more secure if they can maintain {\em some} state across boots --
in particular, Tor normally maintains the same {\em entry relay}
for several months --
and may increase this period further~\cite{dingledine13guards, elahi12guards}.
Users whose pseudonymous actions are linkable
(e.g., via Alice's twitter feed)
are inherently vulnerable to long-term intersection attacks,
which the attacker can exploit far more rapidly
if Tor chooses new entry relays frequently~\cite{johnson13users}.
Thus, if Alice uses a pure amnesiac system to post to her Twitter feed,
Tor is forced to choose a new entry relay each time she boots,
greatly increasing her vulnerability to intersection attacks.

Both to offer convenience by enabling users to maintain content
such as bookmarks, usernames and passwords, and application preferences,
and to ensure that related anonymizer state is also preserved for security,
\winon supports {\em quasi-persistent data}.
Quasi-persistent data resides on the machine only when actively in use.
When not in use, an encrypted copy of the data is migrated
to another storage device -- either to another local partition or USB drive,
or to the cloud,
akin to CleanOS~\cite{tang12cleanos}.
\winon allows users to store
information and data accumulated during a pseudonym session
anonymously into the cloud, thus retaining pseudonym information
while leaving no potentially ``suspicious'' state (even encrypted)
on local devices that might be inspected or confiscated.

\winon supports three different nym usage models:
amnesiac/ephemeral, persistent, and pre-configured.
The latter two both make use of quasi-persistent data,
but with different intent.
In {\em persistent} mode,
\winon updates the nym's stored state after each session,
presenting a familiar and convenient state management model
but increase risk that the effects of a stain or other exploit attack
in one browsing session will persist for the lifetime of the nym.
In {\em pre-configured} mode,
a user boots a nym once,
configures it with appropriate software, settings, bookmarks,
pseudonymous account credentials,
and any other useful state,
then directs \winon to {\em snapshot} the nym.
Each subsequent use of this nym then starts from this snapshot,
never updating the stored nym state unless the user
explicitly requests another snapshot.
Thus, a malware infection affecting one browsing session
will be scrubbed at the user's next session,
and even if the nym's state is eventually compromised,
the attacker obtains no record of the user's client-side activities
using the nym.


\paragraph{Workflow:}
In a typical workflow,
\winon on boot presents the user with a Nym Manager,
offering options to
{\bf start a fresh nym} or {\bf load an existing nym}.
On first use,
the user selects {\bf start a fresh nym}.
Each new nym begins with a writable virtual disk image
in a standard, pristine state.
When done browsing,
if the user opts to store his nym,
he returns to the Nym Manager and selects {\bf store nym}.
The user enters a name for the nym, a password to encrypt it with,
and an indication of a cloud service on which to store the nym.
The Nym Manager navigates the user to the cloud service,
using the CommVM's anonymizer to protect this connection,
and prompts the user to login to the cloud service.
In the background,
the nym manager pauses the nym's AnonVM and CommVM,
syncs their file systems,
compresses and encrypts their temporary file system disk images,
resumes the VMs,
and uploads the contents through the nym's CommVM.
The nym manager notifies the user
once the nym has been saved,
after which the user may close the nym or turn off the computer.

Later the user returns to \winon
and selects {\bf load an existing nym}.
While the Nym Manager prompts the user to select the cloud service
hosting the nym,
\winon starts an {\em ephemeral} nym
for the purpose of gathering the nym's state anonymously from the
selected service.
As before, the nym manager directs the user
to the cloud service's login page,
and then prompts the user for the name of the nym
and the decryption password.
In the background,
the nym manager downloads the nym's state,
and terminates the ephemeral nym used for downloading.
The nym manager then proceeds to
decrypt and decompress the loading nym's CommVM and AnonVM images,
and resumes the nym by starting a new set of VMs using these images.
The user may then continue using the nym.

\paragraph{Security Tradeoffs:}
The cloud storage solution has the advantage of
offering plausible deniability to a user
whose devices or USB drives may be inspected or confiscated.
By utilizing free-to-use cloud storage options,
such as DropBox or Google Drive,
a user can create a pseudonymous cloud account
for each pseudonym.
Because all interactions with the cloud storage are anonymized,
the cloud provider learns nothing about the account owner.
Similarly, as pseudonyms store only encrypted data,
cloud providers learn nothing about the pseudonym therein.
\com{	this sounds handwavy; is any integrity check actually done?
If a cloud provider attempted to manipulate the contents of a pseudonym,
then by using an encryption scheme with integrity checks,
the user would detect this and abort using the pseudonym.
}

One subtle downside of the cloud approach is that
the ephemeral CommVM used to load a nym from the cloud
cannot use the nym's ``proper'' CommVM state --
such as Tor entry guards --
because that CommVM state has not been retrieved yet.
While we do not expect it to be easy for an attacker to correlate
the loading of the nym's state through the ephemeral CommVM
with the user's actions using the nym's own stateful CommVM,
a sufficiently powerful and all-seeing attacker might in principle do so,
making this ephemeral CommVM one remaining point of vulnerability
to long-term intersection attacks.
One solution is for the user simply
to use local storage instead of the cloud.
Another possible solution we are exploring
is to {\em seed} critical CommVM state such as entry guard choices
using a deterministic hash based on the nym's storage location and password,
ensuring that the same seed (and hence same entry guard choices)
get used even by the ephemeral CommVM that load the nym.

\subsection{Sanitized File Transfers}

Users may want to
distribute content from non-anonymous sources --
e.g., Bob wants to post pictures he took on his digital camera
of the day's protests in Tyrannimen Square.
Na\"ively posting such files are risky,
as they may leak the user's identity via
GPS coordinates in EXIF metadata for example%
~\cite{byers04leakage,castiglione07advantages,oakes12hacking}.

To mitigate such risks,
\winon never gives a \nymvm direct access
to files on the client machine's installed OS.
Instead, \winon delegates this responsibility
to a dedicated, non-networked {\em sanitation VM} or SaniVM.
Within this SaniVM,
the user can access files on the installed OS
and transfer files between nyms,
via {\em scrubbing} tools that assist the user
in transferring files safely.

Upon boot, \winon searches the computer for file systems
unrelated to \winon and mounts them in the SaniVM.
Within this SaniVM
the user can browse through all files on the computer
and select files for transfer into a \nymvm.
Prior to making any data accessible in the \nymvm, however,
the SaniVM launches a suite of scrubbing tools that
inspect the files to be transferred,
attempt to identify potential risks such as hidden metadata
or visible faces in photos,
present the user a list of these files and potential risks,
and offer to apply appropriate scrubbing transformations
under control of the user
to remove potentially identifying personal information.

\winon creates a unique directory within the SaniVM for each nym.
The SaniVM detects when the user moves files into this directory
and launches the scrubbing workflow.
Once scrubbing completes,
the SaniVM finally copies the file into a directory
visible to the appropriate nym's AnonVM.

The SaniVM's scrubbing process builds on
the Metadata Anonymization Toolkit (MAT)~\cite{voisin13mat},
but \winon adds atop MAT both additional anonymization methods
and a more user-friendly workflow
incorporating automated risk analysis and identification,
enabling the user to select among alternative transformations,
which might be seen as corresponding to different ``paranoia levels.''
With images, for example,
the user might choose any combination of:
(a) scrub EXIF or other metadata,
(b) blur any detectable faces using OpenCV~\cite{opencv}, and/or
(c) reduce the resolution and add noise in attempt to disrupt
any watermarks the image might contain unbeknownst to the user.
With PDF or DOC files,
the user can similarly scrub metadata,
but also has the option to reconstruct the document completely
as a series of bitmaps,
effectively scrubbing any nonvisual information
that might be concealed (accidentally or intentionally)
in document's complex text or vector graphics structures.

While \winon builds on a wealth of existing techniques
to strip files of potentially identifying material~\cite{
	aura06scanning,bier09rules,sweeney96replacing,voisin13mat},
clearly no scrubbing suite can be perfect.
Developers continuously create new file types,
and add extensions to existing file types,
which might conceal identifying information.
Adversaries can also find improved ways
to exploit existing file types.
Nevertheless, by designing personal information detection, analysis,
and scrubbing into \winon's {\em only} cross-nym file transfer path,
we hope \winon's architecture will ensure that users are at least
made aware of the risks and offered choices that
increase safety in common-case situations.

\subsection{Installed OS as a Nym}

Even if a user boots \winon from USB,
he likely already has a conventional OS installed on the machine,
which he may use for common non-sensitive, non-anonymous activities.
This installed OS is likely to have network-related state
such as WiFi passwords or VPN software
the user regularly employs to access local networks.
To reduce \winon's deployment burden
and network configuration effort required on startup,
\winon can boot the machine's installed OS
in a (non-anonymous) \nymvm,
and leverage its existing state to sign onto relevant WiFi LANs,
or enable the user to find (or create) files on his installed system
that he may wish to transfer to nyms via the SaniVM,
using already-familiar applications on the installed OS.
\winon can currently boot several versions of Windows and Linux in this way.

The three keys challenges for installed OS nyms
that differ from traditional nyms are
booting the OS in a VM,
addressing persistency,
and specifying network configuration parameters.
While Linux usually boots without issue,
booting in a VM a Windows instance installed on the ``bare metal''
can trigger device driver complaints.
We found that a standard repair process typically addresses this problem,
however.

A user's installed OS may of course be compromised
with malware or censorware~\cite{knockel11three}
that may attempt to track or fingerprint the user.
To maximize the safety of booting the installed OS,
\winon treats the machine's hard disk as read-only
and boots the installed OS into a copy-on-write virtual disk,
so that no changes the installed OS makes while running under \winon
ever persist on the physical disk it was booted from.
This design:
(a) ensures that the user will not need to run a repair process
{\em again} when switching back to the installed OS on the bare metal;
(b) ensures that any other unexpected glitches caused
by booting the installed OS in a VM cannot unexpectedly break
the installed OS image on the underlying disk;
(c) avoids leaving any history indicating \winon's use on the local disk,
offer the user plausible deniability.

If a user needs persistence,
he may explicitly allow writes back to the physical disk,
or store his copy-on-write COW disk as quasi-persistent data.
If he subsequently starts his installed OS outside of \winon, however,
it may need to be repaired again, as in the case of Windows.
Further, attempting to use the quasi-persistent COW disk
after the underlying disk has changed can lead to inconsistency or corruption.
Thus, we consider it safest to treat the installed OS as read-only,
and leave exploration of more sophisticated alternatives to future work.

%% file: impl.tex
\section{Prototype Implementation}
\label{sec:impl}

Our prototype \winon systems implements
the architecture discussed in Section~\ref{sec:arch}
including support for various anonymizers,
circumvention tools, and sanitization techniques.
\winon uses the Ubuntu 14.04 64-bit Linux distribution
and QEMU/KVM for running all \nymvms,
besides Windows Host OS nyms.
We have yet to consolidate nym activity to a single interface;
instead we use the graphical user interface of each AnonVM
to host the nym's Web browser.
The Chromium Web browser was chosen
in order to support circumvention software,
specifically StegoTorus~\cite{weinberg12stegotorus}.
We have released \winon source code,
packages that build a fresh \winon disk image,
and \winon images at {\em blinded url}.

\subsection{Anonymizers and Circumvention Tools}

\winon has the necessary configuration to support
anonymizers, circumvention tools, and other communication tools
that use either a SOCKS~\cite{rfc1928}
or virtual network interfaces, such as a tap device.
The entire run-time configuration for these tools resides within the CommVM,
running completely transparent to the AnonVM.
While Tor does not support UDP redirection,
it has a built-in DNS server.
Dissent, on the other hand, does have support for UDP redirection.
For tools that support neither,
\winon would need to convert UDP-based DNS requests to TCP
before transmitting them over the communication tool.

Currently, we have tested the following tools within \winon:
Tor~\cite{dingledine04tor}, Dissent~\cite{wolinsky12scalable},
our own implementation of SWEET~\cite{houmansadr12sweet},
and an incognito mode.
Tor has a built-in DNS server,
while both Dissent and SWEET support UDP based proxying.
Our incognito mode makes use of Linux' IPTables masquerade mode
in order to provide a NAT interface into the Internet.

\subsection{Virtual Machine Management}

For virtualization,
\winon primarily depends on KVM~\cite{KVM},
a virtualization solution built directly into the Linux kernel.
KVM builds upon and recently merged with QEMU~\cite{bellard05qemu}
and takes advantage of hardware virtualization, where available.

The hypervisor configures the network on the AnonVM
to talk directly to the CommVM via a UDP port,
effectively setting a virtual wire connecting the two machines
or a host-only network.
Because that UDP sockets run in the hypervisor,
only applications in the hypervisor can access it.
The CommVM connects to the Internet
by way of KVM user-mode NAT.

\winon configures the VM to reduce the ability
for an adversary to fingerprint a VM.
Each independent set of AnonVMs and CommVMs
have the same Ethernet and IP addresses.
The resolution within an AnonVM is consistently set to 1024x768,
albeit that is configurable upwards and downwards,
we want \winon to run the same on every machine.
Each VM has only a single CPU listed in /proc/cpuinfo as a QEMU Virtual CPU.
The VM has 256 MB writable storage and 256 MB RAM,
both of these consume the host's RAM.

\winon enables KSM or kernel samepage merging.
KSM is a memory-saving de-duplication feature
that scans pages and merges when applicable.
Because all \winon VMs and the hypervisor
use the same disk image and hence applications,
\winon can save a bit of RAM through the use of KSM,
as we show in our evaluations.

\winon stacks the file systems together using OverlayFS,
a union file system built directly into the Linux kernel.
Each VM has three file systems:
1) the base image,
2) a configuration image,
3) and a writable image.
The hypervisor and VMs all share a common base image,
this is the OS installed on the USB stick.
The configuration image masks configuration files
on the base image to enable AnonVM, CommVM, or SaniVM functionality.
The writable image can either be tossed at the end of a session
or stored in the cloud for quasi-persistent data stores.

Many modern virtual machine management tools support
loading a real path within the hosts file system onto a guest.
KVM makes use of VirtFS~\cite{jujjuri10virtfs}.
Within \winon each of the different configuration file systems
exists as paths within the disk image.
When starting the pseudonym VMs,
\winon attaches the appropriate path to the VM
as a VirtFS.

\subsection{Sanitized File Transfers}

The SaniVM hosts a multipurpose scrubbing tool that we designed.
The scrubbing tool runs in two modes,
the first takes advantage of MAT~\cite{voisin13mat},
the Metadata Anonymisation Toolkit.
The second mode converts the document into a series of images,
effectively loading the document into a proper viewer,
taking one or more screen shots,
and then assembling the images together.
Both tools strip away metadata;
however, our extension does so by requiring only a viewing tool
and not a tool that has explicit knowledge about what fields
should be stripped.
Of course, a malicious entity may embed visible content
that neither stripper can remove.

After scrubbing, the SaniVM moves it into a shared folder with the hypervisor.
The hypervisor, then in turn, moves it into a shared folder
with the specific AnonVM.
KVM includes a shared folder technology called VirtFS~\cite{jujjuri10virtfs}.

\com{
Unfortunately VirtFS leaks some information about the host,
namely the total free space,
and in the case of the AnonVM, this can be used to fingerprint the host.
In order to secure against this,
we will need to modify VirtFS to offer
a fixed total, used, and available storage.
}

%% file: eval.tex
\section{Evaluation}
\label{sec:eval}

\subsection{Validating the System}

We validate the \winon prototype
using KVM and nested virtualization
This process made it easy to verify the state of the system
and inspect for potential information leaks.
To check for leaks,
we connected the \winon hypervisor to a virtual network interface
that tunneled traffic to a NAT running on the host.
On the host device,
we ran Wireshark and inspected traffic entering and exiting an idle \winon client.
The \winon hypervisor emitted only traffic for DHCP and anonymizer traffic,
while the AnonVM transmitted no traffic.

We also started many pseudonyms simultaneously
in order to verify the restricted communication model.
We attempted to transmit Ethernet and IP packets
from one AnonVM as well as one CommVM to the local network,
other AnonVMs and CommVMs,
as well as the hypervisor.
All attempts failed with a no-response,
as if the host did not exist.
The AnonVM can only communicate with a functional CommVM
and the CommVM could only communicate with the Internet
not local intranets.

Beyond internal validation,
\winon has been regularly scrutinized
for over 2 years by an independent red-team.

\subsection{Concurrent Nym Usage}

As users explore the new functionality provided by \winon,
there will be a natural increase in pseudonym usage.
Each additional pseudonym costs RAM
as well as induces network and CPU overhead on other pseudonyms.
In this section,
we investigate these overheads in a series of experiments
using an Intel I7 quad core desktop with hardware virtualization extensions
and 16 GB of RAM.
The desktop connects to a test Tor deployment
running on the DeterLab testbed
that eventually reaches the real Internet.
The network connection between the DeterLab testbed~\cite{deterlab},
has a round trip latency of 80ms and throughput
and has been rate limited to 10 Mbit/s through the Linux tool qc,
the DeterLab testbed has no additional delays or bandwidth constraints.
While we could use the real Tor network,
our evaluations focus on the overheads of \winon
and not noise introduced by the dynamic and complex nature of Tor.
To analyze overheads,
the VMs used two different memory configurations.
Our CPU benchmark, Peacemaker,
demanded around 1 GB of RAM,
whereas,
bandwidth and regular Web access
required only 384 MB of RAM.
In all tests,
we allocated 16 MB disk space and 128 MB RAM to each CommVM
and 128 MB disk space to each AnonVM.
The host allocates disk and RAM from its own stash of RAM,
thus limiting the maximum number of nyms.

\begin{figure*}[t]
  \centering
  \begin{minipage}{.32\linewidth}
    \centering
    \includegraphics{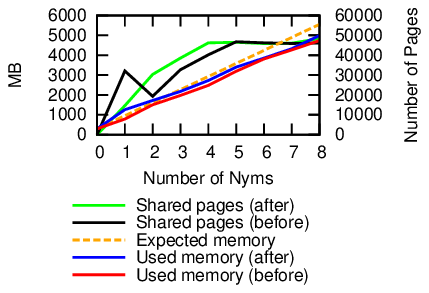}
    \captionof{figure}{RAM usage and shared pages with varying number of pseudonyms
    before and after the new pseudonym becomes active.
    The dash line represents the estimated cost in RAM per-pseudonym.}
    \label{fig:mem}
  \end{minipage}%
  \hspace{2mm}%
  \begin{minipage}{.32\linewidth}
    \centering
    \includegraphics{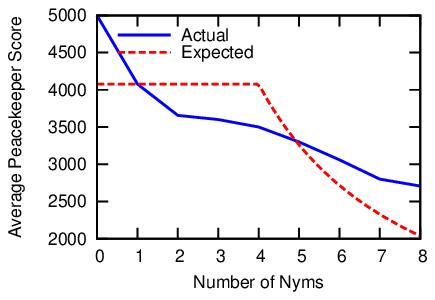}
    \captionof{figure}{Accumulated values for parallel running instances of Peacekeeper
    running in independent pseudonyms.
    0 represents the evaluation when run directly on the host.}
    \label{fig:cpu}
  \end{minipage}%
  \hspace{2mm}%
  \begin{minipage}{.32\linewidth}
    \centering
    \includegraphics{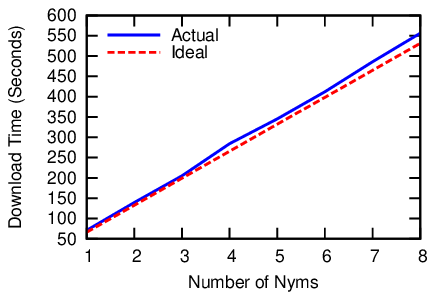}
    \captionof{figure}{Time to download the Linux kernel
    with many nyms downloading in parallel.
    0 represents the evaluation when run directly on the host.\vspace{5ex}}
    \label{fig:net}
  \end{minipage}
\end{figure*}

To evaluate memory per pseudonym,
we launched a series of pseudonyms in succession.
Upon loading a pseudonym,
we checked the current used memory and
kernel samepage merging (KSM) shared pages.
We then interacted with a website
and again noted the used memory and shared pages.
At which point, we loaded another pseudonym and repeated
producing 8 different nyms.
We accessed the following websites in order:
Gmail, Twitter, Youtube, Tor Blog, BBC, Facebook, Slashdot, and ESPN.
Where applicable, we signed into Web sites
and simulated some typical user behaviors,
such as reading the latest news.
Our results, Figure~\ref{fig:mem},
show that KVM obtains most of the requested memory
for a VM at VM initialization and not during run time.
We also see that as more VMs are allocated,
KSM manages to reduces overall memory usage
resulting in over 5\% saving at 8 nyms.

To evaluate CPU overhead,
we ran a Javascript benchmark
called Peacekeeper~\cite{peacekeeper}
in several pseudonyms, simultaneously.
Unfortunately certain experiments with Peacekeeper
consume too much memory causing Chrome to crash,
therefore we had to increase the RAM allocated to the AnonVM
for this evaluation.
We ran the evaluation with up to 8 pseudonyms
and present the results in Figure~\ref{fig:cpu}.
In this graph, 0 represents the system running in native mode.
Virtualization incurs about a 20\% overhead.
When running Peacemaker in parallel,
the actual performance outperforms the expected results,
based upon the single nyms performance when run
multiple times perfectly in parallel with other nyms.
These results indicate that CPU performance overheads,
while apparent, should not be a significant impediment
to \winon scalability.

Each additional nym uses its own instance of an anonymizer
incurring  additional bandwidth overhead due to control messages.
In this evaluation,
we download the current Linux kernel version 3.14.2,
from a server running within DeterLab
in order to guarantee the 10 Mbit download rate.
We varied the number of parallel downloading nyms
and present the results in Figure~\ref{fig:net}.
As we scale the number of nyms, the performance remains relatively linear,
indicating that Tor, the anonymizer in the CommVM, has a fixed cost,
approximately 12\% overhead.
Performance on the real Tor network may differ significantly.

\subsection{Pseudonym Storage}

\begin{figure*}[t]
  \centering
  \begin{minipage}{.48\linewidth}
    \centering
    \includegraphics{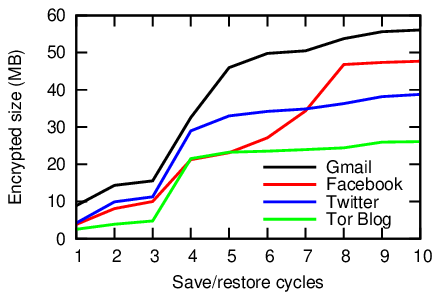}
    \captionof{figure}{Sizes of quasi-persistent pseudonym data across save/restore cycles.}
    \label{fig:size}
  \end{minipage}%
  \hspace{2mm}%
  \begin{minipage}{.48\linewidth}
    \centering
    \includegraphics{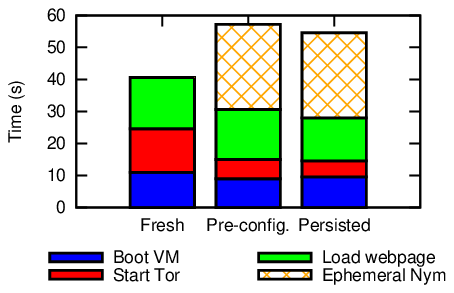}
    \captionof{figure}{Average startup time by phase for each initial configuration.}
    \label{fig:nym_boot}
  \end{minipage}
\end{figure*}

To evaluate the storage requirements for quasi-persistent pseudonyms,
we monitored the size on disk of the nym state
across ten save/restore cycles
over the course of three days.
Both the AnonVM and CommVM were equipped with 256 MB disks.
We performed the experiment with four different nyms
each visiting a different site:
Twitter, Facebook, Gmail, or the Tor Blog.
We began by launching a new pseudonym,
visit the website, sign-in when applicable,
and configure the browser to remember login information.
We then closed the browser and saved
the pseudonym to cloud storage.
For all subsequent measurements,
we restored the nym from cloud storage
and launched the browser,
triggering a fetch of any new site updates.
After the page finished loading,
we closed the browser, and, in the case of persistent nyms,
saved the pseudonym back to cloud storage.
For each upload we recorded the size on disk of the
archived, encrypted pseudonym image.

We present the results of our experiment in Figure~\ref{fig:size}.
Persistent nyms do grow over time
with the AnonVM content accounting for $85\%$ of the pseudonym size,
though much of that is dominated by contents in Chromium cache,
which could have been configured to be smaller
than the default of 83 MB.
Effectively a single save cycle represents
usage similar to a pre-configured nym,
which tends to be small in the order of megabytes.

\subsection{Pseudonym Restoration}

To evaluate the overhead incurred by starting a fresh nym
and restoring a quasi-persistent nym,
we compared startup times for the three different nym usage models:
ephemeral, pre-configured, and persistent.
For each configuration,
we visited the Twitter website and retrieved updates.
Upon finishing the page load,
we closed the browser and,
in the first two models, discarded all changes.
For persistent nyms, we saved all changes back to the persistent state.
We further divided pseudonym startup time into three phases:
AnonVM boot time,
Tor startup time,
and webpage load time.
For quasi-persistent nyms, we include the time it takes
to start an ephemeral nym,
download the state from the cloud,
decrypt it,
and prepare a new nym encapsulated as ``Ephemeral Nym''
We timed five executions from each initial
configuration and present the averaged results in Figure~\ref{fig:nym_boot}.
Quasi-persistent nyms consistently outperform ephemeral nyms
due to stored Tor state;
however, they require a one-time use ephemeral nym to download the data
as well as user interaction necessary for authentication
and nym selection, which was not measured in this evaluation.
In practice, we expect user-interaction to play a significant role
in nym boot times.

\subsection{Installed OS as a Nym}

Running an installed OS as a nym requires both user interaction
and computation resources.
User interaction comes in the form of a user executing commands
to repair the OS to support the change of hardware.
The repair process analyzes the OS state and performs some reconfiguration.
While hard to separate the human process from the automated process,
this evaluation takes a look at both the time required to perform this process
and the resulting impact on memory.
Finally, after repairing the OS,
the operating system can be booted as a nym.
We present the results in Table~\ref{tab:hostos}.
Memory sizes suggest that the repair process is quite invasive,
suggesting that perhaps there may be a more intelligent means to
automating this process.
Ideally we could automate the repairing of the OS in the background,
merging the time a repair takes
with the booting process of \winon and the associated CommVM.

\begin{table}[h]
\centering
\begin{small}
\begin{tabular}{l|ccc}
  & Repair (S) & Boot (S) & Size (MB) \\ \hline
  Vista & 133.7 & 37.7 & 4.9 \\
  7 & 129.3 & 34.3 & 4.5 \\
  8 & 157.0 & 58.7 & 14 \\
\end{tabular}
\end{small}
\caption{Time and memory costs of using various versions of Windows
as a nym in \winon.}
\label{tab:hostos}
\end{table}

%% file: rel.tex
\section{Related Work}
\label{sec:rel}

The \winon project bears resemblance
to other safety-focused bootable media projects.
Tails~\cite{tails} uses external bootable devices
which have been preconfigured with Tor,
the Tor version of Firefox,
and other utilities for privacy protection.
Like \winon,
Tails supports persistent state;
however, Tails stores that state on the same USB as Tails.
Even if the data is encrypted,
a sufficiently powerful adversary could likely coerce
the key from its owner.
\winon offers deniability
by storing persistent state anonymously to the cloud.
Because of the large community surrounding Tails,
we view \winon as a development platform for research ideas
that eventually will be integrated into Tails,
namely, structural protection such as \nymvms and quasi-persistent data.

Like \winon, Whonix~\cite{whonix} separates the user environment
from the communication tool eliminating accidental information leaks
due to faulty component configurations.
Unlike \winon, Whonix installs a pair of virtual machines
on the users installed operating system.
Unlike \winon, Whonix cannot defend against hardware fingerprinting,
confiscation, or correlation attacks.

Similar to how \winon isolates user data into a single environment, the SaniVM,
the US military~\cite{karger05multilevel} and NSA~\cite{meushaw00nettop}
have made similar efforts to separate classified and unclassified information.
In fact, the US military's approach to separation
entitled ``secure shared file store options''
has a similar construction to \winon's use of VirtFS
without requiring changes to the hypervisor.

\winon solves an important problem in the domain of scrubbing,
negligence in their use~\cite{byers04leakage},
by forcing their use as part of an OS primitive
Earlier work proposes similar extensions for
networks~\cite{ioannidis06privacy}.
\winon builds on MAT~\cite{voisin13mat},
but there are many other tools out there
and metadata scrubbing may be insufficient.
Even intentionally blinded conference papers
leak personally identifiable information~\cite{aura06scanning}.
Bier~\cite{bier09rules} shows that removing keywords is insufficient
and need semantic analysis,
in effect, nothing will be perfect.
Data may be hidden by steganography
and there is not much a scrubber can do to prevent
active steganography~\cite{castiglione07advantages}.

As an alternative to selecting files and folders to scrub,
\winon could employ concepts introduced by
User-Driven Access Control~\cite{roesner12user}.
In this model,
a user could grant access to certain folders and files on the host
to a specific nym.
\winon could then delay scrubbing of files
until the files have been accessed from within the nym.

The notion of pseudonyms long predates anonymous digital communication.
Recent work from Han et al.~\cite{han13expressive} explores the notion
of a pseudonym browsing mode.
Their work focuses on a network adversary
who links users primarily by IP addresses.
To mitigate the effectiveness of this adversary,
each pseudonym runs in the same OS but
within different browser profiles and IPv6 addresses.
Their approach requires infrastructure changes to deployed IPv6 routers
in order to support multiple IPv6 addresses at the same site
that could not easily be correlated,
effectively creating a one hop proxy or a VPN.
They have also implemented a plugin
to address adversaries similar to Panopticlick~\cite{panopticlick};
however, this is an arms race
and the plugin will need to be maintained in order to ensure
that future fields do not leak information.
\winon's structural approach to homogeneity
offers a future proof architecture
that remains immunity to these types of attacks.

On the other end of the spectrum,
there has been considerable work in using OS sandboxing
to enhance isolation and resistance
to compromised browsers
and applications, such as BOS~\cite{cox06bos}, IBOS~\cite{tang10trust},
and Atlantis~\cite{mickens11atlantis}.
These approaches separate each web page instance
into unique processes
that have access to a limited API
intended for web browsing only.
However, this work has yet to address the need
for pseudonymity and anonymity.
Similar efforts have been made to offer separation
on content in general and not just Web interaction~\cite{moshchuk12content},
a feature \winon also seamlessly offers.
\winon currently uses virtual machines to offer sandboxing,
but architecturally \winon could in principle
sandboxing could be offered by lightweight process-level sandboxes~\cite{moshchuk12content},
software fault isolation~\cite{wahbe93efficient,yee09native},
enhanced browser-based sandboxes%
~\cite{tang10trust, mickens11atlantis, ingram12treehouse, cox06bos},
virtual machines~\cite{cox06bos},
or even hosting \winon instances in the cloud~\cite{martignoni12cloud}.

\winon could be extended in many directions.
A nym only isolates identities,
but could benefit from approaches
that isolate and protect sensitive data in a browser,
such as Configurable Origin Policies~\cite{cao13redefining},
TaintDroid~\cite{enck10taintdroid},
and Crypton-Kernel~\cite{dong13protecting}.
While \winon might isolate a key logger,
ScreenPass~\cite{liu13screenpass} could offer
\winon a means to secure password entry
to avoid spoofing attacks by providing a trusted password entry keyboard.

%% file: disc.tex
\section{Discussion and Future Work}
\label{sec:discussion}

\paragraph{The Enemy Within:}
The \winon model depends on the sterility of both hardware and software.
A malicious party could install malware
into the hypervisor prior to distribution
or in the firmware of WiFi devices prior to a party
in order to compromise a user's anonymity.
Using trusted platform modules could potentially ensure the running software
and firmware; however,
all is for naught if the hardware vendor
has conspired with the adversary.

\paragraph{Lack of Perfect Homogeneity:}
Even while using virtualization and the same set of software,
there still exists the possibility for differences between users.
An adversary could execute a particularly CPU intensive application,
such as a Javascript application that computes a million digits of PI,
and use the timings to produce a fingerprint for that user.
Also, all users cannot necessarily be in the same location,
and hence if there is a single Tor user in Tyrannistan,
the government-owned ISP could easily
determine the responsible party for any Tyrannistani traffic
related to Tor.

\com{
\paragraph{Towards Deterministic Nym Behavior:}
}

\com{
\paragraph{Self-Deanonymizing Users:}
A reckless user can easily de-anonymize himself in most anonymity systems,
while \winon goes to great lengths to prevent such accidents,
there exist some limitations.
In \winon, for example, if a user shares personal information
while signing an anonymous petition,
by providing a personal e-mail account,
\winon like most other systems would be unable
to prevent this type of accident.
However, because of nym-browse,
a user need not worry about 
correlation attacks wherein they de-anonymize a single nym instance.
Specifically, if a user were to browse their Facebook or
personal webmail account in one tab
and access sensitive material in another,
the adversary would be unable to link this material together.
In systems like Tails and Whonix,
such behavior might instantly de-anonymize the user.
}

\paragraph{Long Term Intersection Attacks:}
\winon mitigates intersection attacks by reducing a user's fingerprint;
however, a fingerprint can be developed even if the user employs
amnesia simply by the set of websites he visits
or the accounts used on those sites.
An adversary performs an intersection attack~\cite{raymond00traffic}
by tracking the online set of participants
and discovering a set of linkable, yet anonymous messages.
The adversary constructs an intersection of users
that were online at the same time as those linkable messages.
With sufficiently many number of messages,
the adversary will be able to discover the owner of the linkable messages.
To enhance \winon's ability to resist intersection attacks,
we plan to integrate Buddies~\cite{wolinsky13buddies}.
Buddies offers users anonymity metrics
and safe guards a user from falling below a desirable anonymity threshold.

\paragraph{Concealing Network Identity:}
Network identity proves difficult
even in the \winon context.
Network fingerprinting comes in many forms
from operating system interfaces to NIC devices%
~\cite{pang07fingerprinting},
drivers~\cite{franklin06fingerprinting},
MAC addresses,
and even the hardware characteristics of devices%
~\cite{brik08fingerprinting}.
Some of these attacks are avoidable
using a common device with a standardized driver,
a volatile management framework for the device,
and randomized MAC addresses.

For well-equipped adversaries,
this approach is insufficient.
Brik et al.~\cite{brik08fingerprinting} determined
that even devices from the same manufacturer
with sequential serial numbers could be fingerprinted
due to the errors in the signal.
Since we envision
that \winon may be used in moderately well-organized groups,
we posit
that users could organize WiFi device exchange parties,
or WiFi social mixes,
akin to Richard Stallman's ``Charlie Card''
swapping parties~\cite{sedgwick08charlie}
to elude RFID-based fingerprinting.
During these parties,
each member would place their cards in a box.
After collecting all members' cards,
each member would randomly select one
without knowing which had been taken and which were left.
Users might have many such parallel exchanges,
so that a user could have several WiFi cards at a time.
In addition, individuals could use an
active antenna to ambiguate their location
as well also to change the physical properties of the device's transmissions.

\com{
  - Jiang~\cite{jiang07preserving}: Discussion of
        open problems in network and physical-layer anonymity.
        Present protocol for refreshing MAC address. 
        Use ``silent periods'' (user forced offline) to hinder
        location tracking of user.
  - He~\cite{he04quest}: Using a ``blind signature'' authentication
        protocol to authenticate to a network without being linkable
        across sessions.
}

%% file: conc.tex
\section{Conclusion}
\label{sec:conc}

\winon offers novel structural solutions
for managing online identities or pseudonyms.
In contrast to existing system solutions
for anonymity and pseudonymity,
\winon provides completely independent state
for each of the user's identities.
\winon, however, does not subsume or replace
the need for other techniques
or hardened systems.
We believe \winon offers a useful platform
for researching Web browsing pseudonymity
that should eventually be incorporated into Tails
and similarly hardened systems.

%% file: main.bbl
\begin{thebibliography}{10}

\bibitem{keepass}
{KeePass}.
\newblock \url{http://keepass.info/}.

\bibitem{opencv}
{OpenCV}.
\newblock \url{http://opencv.org}.

\bibitem{aggrawal10private}
G.~Aggrawal, E.~Bursztein, C.~Jackson, and D.~Boneh.
\newblock An analysis of private browsing modes in modern browsers.
\newblock In {\em Usenix Security Symposium}, 2010.

\bibitem{aura06scanning}
T.~Aura, T.~A. Kuhn, and M.~Roe.
\newblock Scanning electronic documents for personally identifiable
  information.
\newblock In {\em \bibconf[5th]{WPES}{ACM workshop on Privacy in electronic
  society}}, pages 41--50, New York, NY, USA, 2006. ACM.

\bibitem{bellard05qemu}
F.~Bellard.
\newblock {QEMU}, a fast and portable dynamic translator, Apr. 2005.

\bibitem{bier09rules}
E.~Bier, R.~Chow, P.~Golle, T.~King, and J.~Staddon.
\newblock The rules of redaction: Identify, protect, review (and repeat).
\newblock {\em Security Privacy, IEEE}, 7(6):46 --53, nov.-dec. 2009.

\bibitem{brik08fingerprinting}
V.~Brik, S.~Banerjee, M.~Gruteser, and S.~Oh.
\newblock Wireless device identification with radiometric signatures.
\newblock In {\em ACM International Conference on Mobile Computing and
  Networking (MobiCom)}, pages 116--127, 2008.

\bibitem{byers04leakage}
S.~Byers.
\newblock Information leakage caused by hidden data in published documents.
\newblock {\em IEEE Security and Privacy}, 2:23--27, 2004.

\bibitem{cao13redefining}
Y.~Cao, V.~Rastogi, Z.~Li, Y.~Chen, and A.~Moshchuk.
\newblock Redefining web browser principals with a configurable origin policy.
\newblock In {\em \bibconf[43rd]{DSN}{IEEE/IFIP International Conference on
  Dependable Systems and Networks}}, June 2013.

\bibitem{castiglione07advantages}
A.~Castiglione, A.~D. Santis, and C.~Soriente.
\newblock Taking advantages of a disadvantage: Digital forensics and
  steganography using document metadata.
\newblock {\em Journal of Systems and Software}, 80(5):750 -- 764, 2007.

\bibitem{chaum88dining}
D.~Chaum.
\newblock The {Dining Cryptographers} problem: Unconditional sender and
  recipient untraceability.
\newblock {\em Journal of Cryptology}, pages 65--75, Jan. 1988.

\bibitem{cox06bos}
R.~S. Cox, S.~D. Gribble, H.~M. Levy, and J.~G. Hansen.
\newblock A safety-oriented platform for web applications.
\newblock In {\em \bibconf{SP}{IEEE Symposium on Security and Privacy}}, 2006.

\bibitem{deterlab}
{DeterLab} network security testbed, September 2012.
\newblock \url{http://isi.deterlab.net/}.

\bibitem{dingledine13guards}
R.~Dingledine.
\newblock Improving tor's anonymity by changing guard parameters, 2013.
\newblock
  \url{https://blog.torproject.org/blog/improving-tors-anonymity-changing-guard-parameters}.

\bibitem{dingledine04tor}
R.~Dingledine, N.~Mathewson, and P.~Syverson.
\newblock Tor: the second-generation onion router.
\newblock In {\em 12th USENIX Security Symposium}, Aug. 2004.

\bibitem{dong13protecting}
X.~Dong, Z.~Chen, H.~Siadati, S.~Tople, P.~Saxena, and Z.~Liang.
\newblock Protecting sensitive web content from client-side vulnerabilities
  with {CRYPTONs}.
\newblock In {\em \bibconf[20th]{CCS}{ACM conference on Computer and
  communications security}}, Nov. 2013.

\bibitem{duhigg12how}
C.~Duhigg.
\newblock How companies learn your secrets.
\newblock {\em The New York Times}, Feb. 2012.
\newblock
  \url{http://www.nytimes.com/2012/02/19/magazine/shopping-habits.html}.

\bibitem{dunn12eternal}
A.~M. Dunn, M.~Z. Lee, S.~Jana, S.~Kim, M.~Silberstein, Y.~Xu, V.~Shmatikov,
  and E.~Witchel.
\newblock Eternal sunshine of the spotless machine: Protecting privacy with
  ephemeral channels.
\newblock In {\em \bibconf{OSDI}{USENIX Symposium on Operating Systems Design
  and Implementation}}, 2012.

\bibitem{eckersley10browser}
P.~Eckersley.
\newblock How unique is your web browser?
\newblock In {\em \bibbrev{PETS}{Privacy-Enhancing Technologies Symposium}},
  July 2010.

\bibitem{elahi12guards}
T.~Elahi, K.~Bauer, M.~AlSabah, R.~Dingledine, and I.~Goldberg.
\newblock Changing of the guards: A framework for understanding and improving
  entry guard selection in tor.
\newblock In {\em \bibconf{WPES}{Workshop on Privacy in the Electronic
  Society}}, 2012.

\bibitem{enck10taintdroid}
W.~Enck, P.~Gilbert, B.-G. Chun, L.~P. Cox, J.~Jung, P.~McDaniel, and A.~N.
  Sheth.
\newblock Taintdroid: An information-flow tracking system for realtime privacy
  monitoring on smartphones.
\newblock In {\em \bibconf{OSDI}{USENIX Conference on Operating Systems Design
  and Implementation}}, 2010.

\bibitem{fleischer12tor}
G.~Fleischer.
\newblock Attacking tor at the application layer.
\newblock
  \url{http://ww.defcon.org/images/defcon-17/dc-17-presentations/defcon-17-gregory_fleischer-attacking_tor.pdf},
  July 2009.

\bibitem{panopticlick}
E.~F. Foundation.
\newblock \url{https://panopticlick.eff.org/}, Oct 2013.

\bibitem{franklin06fingerprinting}
J.~Franklin, D.~McCoy, P.~Tabriz, V.~Neagoe, J.~Van~Randwyk, and D.~Sicker.
\newblock Passive data link layer 802.11 wireless device driver fingerprinting.
\newblock In {\em USENIX Security Symposium}, 2006.

\bibitem{peacekeeper}
Futuremark.
\newblock Peacekeeper -- the universal browser test.
\newblock \url{http://peacekeeper.futuremark.com}, January 2013.

\bibitem{gassend03merkle}
B.~Gassend, G.~Suh, D.~Clarke, M.~van Dijk, and S.~Devadas.
\newblock Caches and hash trees for efficient memory integrity verification.
\newblock In {\em \bibconf{HPCA}{High-Performance Computer Architecture}},
  2003.

\bibitem{goodin13attackers}
D.~Goodin.
\newblock Attackers wield {Firefox} exploit to uncloak anonymous {Tor} users.
\newblock {\em ars technica}, Aug. 2013.

\bibitem{gu11certikos}
L.~Gu, A.~Vaynberg, B.~Ford, Z.~Shao, and D.~Costanzo.
\newblock {CertiKOS}: A certified kernel for secure cloud computing.
\newblock In {\em \bibconf[2nd]{APSys}{ACM SIGOPS Asia-Pacific Workshop on
  Systems}}, July 2011.

\bibitem{han13expressive}
S.~Han, V.~Liu, Q.~Pu, S.~Peter, T.~Anderson, A.~Krishnamurthy, and
  D.~Wetherall.
\newblock Expressive privacy control with pseudonyms.
\newblock In {\em \bibbrev{SIGCOMM}{ACM SIGCOMM}}, Aug. 2013.

\bibitem{hill12target}
K.~Hill.
\newblock How target figured out a teen girl was pregnant before her father
  did.
\newblock {\em Forbes}, Feb. 2012.
\newblock
  \url{http://www.forbes.com/sites/kashmirhill/2012/02/16/how-target-figured-out-a-teen-girl-was-pregnant-before-her-father-did/}.

\bibitem{hill14pregnancy}
K.~Hill.
\newblock You can hide your pregnancy online, but you'll feel like a criminal.
\newblock {\em Forbes}, Apr. 2014.
\newblock
  \url{http://www.forbes.com/sites/kashmirhill/2014/04/29/you-can-hide-your-pregnancy-online-but-youll-feel-like-a-criminal/}.

\bibitem{houmansadr12sweet}
A.~Houmansadr, W.~Zhou, M.~Caesar, and N.~Borisov.
\newblock Sweet: Serving the web by exploiting email tunnels.
\newblock {\em CoRR}, abs/1211.3191, 2012.

\bibitem{howard13democracy}
P.~N. Howard and M.~M. Hussain.
\newblock {\em Democracy's Fourth Wave? Digital Media and the Arab Spring}.
\newblock Oxford University Press, Mar. 2013.

\bibitem{ingram12treehouse}
L.~Ingram and M.~Walfish.
\newblock {TreeHouse}: {JavaScript} sandboxes to help {Web} developers help
  themselves.
\newblock In {\em \bibconf{ATC}{USENIX Annual Technical Conference}}, June
  2012.

\bibitem{ioannidis06privacy}
S.~Ioannidis, S.~Sidiroglou, and A.~D. Keromytis.
\newblock Privacy as an operating system service.
\newblock In {\em Proceedings of the 1st USENIX Workshop on Hot Topics in
  Security}, HOTSEC'06, pages 8--8, Berkeley, CA, USA, 2006. USENIX
  Association.

\bibitem{johnson13users}
A.~Johnson, C.~Wacek, R.~Jansen, M.~Sherr, and P.~Syverson.
\newblock Users get routed: Traffic correlation on {Tor} by realistic
  adversaries.
\newblock In {\em \bibconf[20th]{CCS}{ACM Conference on Computer and
  Communications Security}}, Nov. 2013.

\bibitem{jujjuri10virtfs}
V.~Jujjuri, E.~V. Hensbergen, A.~Liguori, and B.~Pulavarty.
\newblock Virtfs--a virtualization aware file system pass-through.
\newblock June 2010.

\bibitem{kamkar10evercookie}
S.~Kamkar.
\newblock evercookie.
\newblock \url{http://http://samy.pl/evercookie/}, oct 2010.

\bibitem{karger05multilevel}
P.~A. Karger.
\newblock Multi-level security requirements for hypervisors.
\newblock In {\em Computer Security Applications Conference, 21st Annual},
  ACSAC '05, Dec. 2005.

\bibitem{kedogan02limits}
D.~Kedogan, D.~Agrawal, and S.~Penz.
\newblock Limits of anonymity in open environments.
\newblock In {\em 5th International Workshop on Information Hiding}, Oct. 2002.

\bibitem{klein09sel4}
G.~Klein et~al.
\newblock {seL4}: formal verification of an {OS} kernel.
\newblock In {\em \bibconf[22nd]{SOSP}{ACM Symposium on Operating System
  Principles}}, Oct. 2009.

\bibitem{knockel11three}
J.~Knockel, J.~R. Crandall, and J.~Saia.
\newblock In {\em Three Researchers, Five Conjectures: An Empirical Analysis of
  {TOM-Skype} Censorship and Surveillance}, Aug. 2011.

\bibitem{leblond13anon}
S.~Le~Blond, D.~Choffnes, W.~Zhou, P.~Druschel, H.~Ballani, and P.~Francis.
\newblock Towards efficient traffic-analysis resistant anonymity networks.
\newblock In {\em {ACM} {SIGCOMM}}, August 2013.

\bibitem{rfc1928}
M.~Leech et~al.
\newblock {SOCKS} protocol, Mar. 1996.
\newblock RFC 1928.

\bibitem{leger13silkroad}
D.~L. Leger.
\newblock How fbi brought down cyer-underworld site silk road, 2013.
\newblock
  \url{http://www.usatoday.com/story/news/nation/2013/10/21/fbi-cracks-silk-road/2984921/}.

\bibitem{lim12clicks}
M.~Lim.
\newblock Clicks, cabs, and coffee houses: Social media and oppositional
  movements in egypt, 2004 -- 2011.
\newblock {\em Journal of Communication}, 62:231--248.

\bibitem{liu13screenpass}
D.~Liu, E.~Cuervo, V.~Pistol, R.~Scudellari, and L.~P. Cox.
\newblock Screenpass: Secure password entry on touchscreen devices.
\newblock In {\em \bibconf[11th]{MobiSys}{International Conference on Mobile
  Systems, Applications, and Services}}, June 2013.

\bibitem{martignoni12cloud}
L.~Martignoni, P.~Poosankam, M.~Zaharia, J.~Han, S.~McCamant, D.~Song,
  V.~Paxson, A.~Perrig, S.~Shenker, and I.~Stoica.
\newblock In {\em \bibconf{ATC}{USENIX Annual Technical Conference}}, June
  2012.

\bibitem{meushaw00nettop}
R.~Meushaw and D.~Simard.
\newblock {NetTop}: Commercial technology in high assurance applications.
\newblock {\em Tech Trend Notes}, 2000.

\bibitem{mickens11atlantis}
J.~Mickens and M.~Dhawan.
\newblock Atlantis: Robust, extensible execution environments for web
  applications.
\newblock In {\em \bibconf[23rd]{SOSP}{ACM Symposium on Operating System
  Principles}}, Oct. 2011.

\bibitem{moshchuk12content}
A.~Moshchuk, H.~J. Wang, and Y.~Liu.
\newblock Content-based isolation: Rethinking isolation policy in modern client
  systems.
\newblock Technical Report MSR-TR-2012-82, Microsoft Research, Aug. 2012.

\bibitem{oakes12hacking}
D.~Oakes.
\newblock Hacking case's body of evidence.
\newblock {\em The Age}, Apr. 2012.

\bibitem{oecd13exploring}
OECD.
\newblock Exploring the economics of personal data, Apr. 2013.
\newblock OECD Digital Economy Papers No. 220.

\bibitem{pang07fingerprinting}
J.~Pang, B.~Greenstein, R.~Gummadi, S.~Seshan, and D.~Wetherall.
\newblock 802.11 user fingerprinting.
\newblock In {\em ACM International Conference on Mobile Computing and
  Networking (MobiCom)}, pages 99--110, 2007.

\bibitem{perry11firefox}
M.~Perry.
\newblock To toggle, or not to toggle: The end of torbutton.
\newblock
  \url{https://blog.torproject.org/blog/toggle-or-not-toggle-end-torbutton},
  May 2011.

\bibitem{mullenize}
W.~Post.
\newblock {GCHQ} report on {`MULLENIZE'} program to `stain' anonymous
  electronic traffic.
\newblock
  \url{http://apps.washingtonpost.com/g/page/world/gchq-report-on-mullenize-program-to-stain-anonymous-electronic-traffic/502/},
  oct 2013.

\bibitem{KVM}
Qumranet.
\newblock Kernel-based virtual machine for linux.
\newblock \url{http://kvm.qumranet.com/kvmwiki}, March 2007.

\bibitem{raymond00traffic}
J.-F. Raymond.
\newblock Traffic analysis: Protocols, attacks, design issues and open
  problems.
\newblock In {\em Workshop on Design Issues in Anonymity and Unobservability},
  pages 10--29, 2000.

\bibitem{risen13nsa}
J.~Risen and L.~Poitras.
\newblock {NSA} report outlined goals for more power.
\newblock {\em The New York Times}, Nov.~22, 2013.

\bibitem{roesner12user}
F.~Roesner, T.~Kohno, A.~Moshchuk, B.~Parno, H.~J. Wang, and C.~Cowan.
\newblock User-driven access control: Rethinking permission granting in modern
  operating systems.
\newblock In {\em IEEE Symposium on Security and Privacy}, May 2012.

\bibitem{schneier13attacking}
B.~Schneier.
\newblock Attacking {Tor}: how the {NSA} targets users' online anonymity.
\newblock {\em The Guardian}, Oct. 2013.

\bibitem{schneier13nsa}
B.~Schneier.
\newblock Nsa surveillance: A guide to staying secure.
\newblock {\em The Guardian}, Sept. 2013.

\bibitem{erratasec12lulzsec}
E.~Security.
\newblock Notes on {Sabu} arrest, 2012.
\newblock \url{http://blog.erratasec.com/2012/03/notes-on-sabu-arrest.html}.

\bibitem{sedgwick08charlie}
J.~Sedgwick.
\newblock The shaggy god.
\newblock \url{http://www.bostonmagazine.com/articles/2008/04/the-shaggy-god/},
  May 2008.

\bibitem{soltani09flash}
A.~Soltani, S.~Canty, Q.~Mayo, L.~Thomas, and C.~J. Hoofnagle.
\newblock Flash cookies and privacy, Aug. 2009.

\bibitem{stein03queers}
E.~Stein.
\newblock Queers anonymous: Lesbians, gay men, free speech, and cyberspace.
\newblock {\em Harvard Civil Rights-Civil Liberties Law Review}, 2003.

\bibitem{sweeney96replacing}
L.~Sweeney.
\newblock Replacing personally-identifying information in medical records, the
  scrub system.
\newblock {\em Journal of the American Medical Informatics Association}, 1996.

\bibitem{tails}
Tails: The amnesic incognito live system, September 2012.
\newblock \url{https://tails.boum.org/}.

\bibitem{tang10trust}
S.~Tang, H.~Mai, and S.~T. King.
\newblock Trust and protection in the illinois browser operating system.
\newblock In {\em \bibconf[9th]{OSDI}{USENIX Symposium on Operating Systems
  Design and Implementation}}, Oct. 2010.

\bibitem{tang12cleanos}
Y.~Tang, P.~Ames, S.~Bhamidipati, A.~Bijlani, R.~Geambasu, and N.~Sarda.
\newblock Cleanos: limiting mobile data exposure with idle eviction.
\newblock In {\em \bibbrev{OSDI}{USENIX Symposium on Operating Systems Design
  and Implementation}}, Oct. 2012.

\bibitem{voisin13mat}
J.~Voisin, C.~Guyeux, and J.~M. Bahi.
\newblock The metadata anonymization toolkit.
\newblock \url{http://arxiv.org/abs/1212.3648}, may 2013.

\bibitem{wahbe93efficient}
R.~Wahbe, S.~Lucco, T.~E. Anderson, and S.~L. Graham.
\newblock Efficient software-based fault isolation.
\newblock {\em ACM SIGOPS Operating Systems Review}, 27(5):203--216, Dec. 1993.

\bibitem{watts13rowling}
R.~Watts.
\newblock {JK Rowling} unmasked as author of acclaimed detective novel.
\newblock {\em The Telegraph}, July~13, 2013.

\bibitem{weinberg12stegotorus}
Z.~Weinberg, J.~Wang, V.~Yegneswaran, L.~Briesemeister, S.~Cheung, F.~Wang, and
  D.~Boneh.
\newblock {StegoTorus}: a camouflage proxy for the {Tor} anonymity system.
\newblock In {\em \bibconf[19th]{CCS}{ACM conference on Computer and
  Communications Security}}, Oct. 2012.

\bibitem{whonix}
Whonix.
\newblock \url{http://sourceforge.net/p/whonix}.

\bibitem{wolinsky12scalable}
D.~Wolinsky, H.~Corrigan-Gibbs, B.~Ford, and A.~Johnson.
\newblock Scalable anonymous group communication in the anytrust model.
\newblock In {\em \bibconf{EuroSec}{European Workshop on System Security}},
  Apr. 2012.

\bibitem{wolinsky13buddies}
D.~I. Wolinsky, E.~Syta, and B.~Ford.
\newblock Hang with your buddies to resist intersection attacks.
\newblock In {\em 20th {ACM} Conference on Computer and Communications Security
  (CCS)}, November 2013.

\bibitem{yee09native}
B.~Yee et~al.
\newblock Native client: A sandbox for portable, untrusted x86 native code.
\newblock In {\em {IEEE} Symposium on Security and Privacy}, May 2009.

\bibitem{zhang11cloud}
Y.~Zhang, A.~Juels, A.~Oprea, and M.~Reiter.
\newblock Homealone: Co-residency detection in the cloud via side-channel
  analysis.
\newblock In {\em \bibconf{IEEE SP}{IEEE Security and Privacy SP}}, 2011.

\bibitem{zhang12cloud}
Y.~Zhang, A.~Juels, M.~K. Reiter, and T.~Ristenpart.
\newblock Cross-vm side channels and their use to extract private keys.
\newblock In {\em \bibconf{CCS}{ACM Conference on Computer and Communications
  Security}}, 2012.

\end{thebibliography}
